\documentclass[a4paper, 11pt]{article}
\usepackage[utf8]{inputenc}
\usepackage[T1]{fontenc}

\usepackage{geometry}
\usepackage{authblk}
\usepackage{graphicx}
\usepackage{setspace}

\usepackage{amsmath}
\usepackage{bm}

\usepackage{float}
\usepackage{booktabs}
\usepackage{longtable}
\usepackage{threeparttable}
\usepackage{makecell}

\usepackage{charter}
\usepackage[bitstream-charter]{mathdesign}

\usepackage[bookmarks=true, bookmarksopen=true, colorlinks, citecolor=blue, urlcolor=blue, linkcolor=blue]{hyperref}
\usepackage[nameinlink,capitalize]{cleveref}
\crefname{chapter}{Chapter}{Chapters}
\crefname{section}{Section}{Sections}
\crefname{subsection}{Section}{Sections}
\crefname{subsubsection}{Section}{Sections}
\crefname{figure}{Figure}{Fig.}
\crefname{table}{Table}{Tab.}
\crefname{equation}{Eq.}{Eqs.}
\crefname{appendix}{Appendix}{Appendices}
\crefname{appsec}{Appendix}{Appendices}
\usepackage{bookmark}
\usepackage{enumitem}

\usepackage[hang,flushmargin]{footmisc} 
\usepackage[yyyymmdd]{datetime}
\usepackage[british]{babel}

\usepackage{float}
\usepackage{bigstrut}
\restylefloat{table}

\usepackage[title,titletoc]{appendix}
\makeatletter

\renewenvironment{appendices}{%
    \begin{oldappendices}%
    \renewcommand{\thefigure}{\ifnum \c@section>\z@ \thesection.\fi\@arabic\c@figure}%
    \@addtoreset{figure}{section}%
    \renewcommand{\thetable}{\ifnum \c@section>\z@ \thesection.\fi\@arabic\c@table}%
    \@addtoreset{table}{section}}{%
    \end{oldappendices}%
}\makeatother

\usepackage{natbib}
\let\natbibcitet\citet
\renewcommand\citet{\bibpunct{(}{)}{,}{a}{,}{,}\natbibcitet}
\let\natbibcitep\citep
\renewcommand\citep{\bibpunct{(}{)}{;}{a}{,}{;}\natbibcitep}
\newcommand{\bi}{\begin{itemize}}
\newcommand{\ei}{\end{itemize}}
\newcommand{\be}{\begin{equation}}
\newcommand{\ee}{\end{equation}}

\long\def\symbolfootnote[#1]#2{\begingroup%
\def\thefootnote{\fnsymbol{footnote}}\footnote[#1]{#2}\endgroup}

\makeatletter
\def\ubar#1{\underline{\sbox\tw@{$#1$}\dp\tw@\z@\box\tw@}}
\def\obar#1{\overline{\sbox\tw@{$#1$}\dp\tw@\z@\box\tw@}}
\makeatother

\newcommand{\N}{\mathcal{N}}

\usepackage{caption}

\makeatletter\let\p@subfigure\thefigure\makeatother

\usepackage[labelformat=simple]{subcaption}

\makeatletter

\def\Autoref#1{%
  \begingroup
  \edef\reserved@a{\cpttrimspaces{#1}}%
  \ifcsndefTF{r@#1}{%
    \xaftercsname{\expandafter\testreftype\@fourthoffive}
      {r@\reserved@a}.\\{#1}%
  }{%
    \ref{#1}%
  }%
  \endgroup
}
\def\testreftype#1.#2\\#3{%
  \ifcsndefTF{#1autorefname}{%
    \def\reserved@a##1##2\@nil{%
      \uppercase{\def\ref@name{##1}}%
      \csn@edef{#1autorefname}{\ref@name##2}%
      \autoref{#3}%
    }%
    \reserved@a#1\@nil
  }{%
    \autoref{#3}%
  }%
}
\makeatother

\title{Implications of macroeconomic volatility in the Euro area}
\author[]{Niko Hauzenberger\thanks{Corresponding author: Niko Hauzenberger, Vienna University of Economics and Business. Address: Welthandelsplatz 1, 1020 Wien, Austria. Email: \href{mailto:niko.hauzenberger@s.wu.ac.at}{niko.hauzenberger@s.wu.ac.at}. The authors thank Florian Huber, who supported us in times of uncertainty.}}
\author[]{Maximilian Böck}
\author[]{Michael Pfarrhofer}
\author[]{Anna Stelzer}
\author[]{Gregor Zens}
\affil[]{%
  \small{WU Vienna University of Economics and Business}}
\date{}


\def\equationautorefname~#1\null{%
  Eq.~(#1)\null
}
\def\equationautorefname~#1\null{
Eq.~(#1)\null
}

\begin{document}
\clearpage\maketitle
\thispagestyle{empty}

\onehalfspacing

\begin{abstract}
\noindent In this paper we estimate a Bayesian vector autoregressive model with factor stochastic volatility in the error term to assess the effects of an uncertainty shock in the Euro area. This allows us to treat macroeconomic uncertainty as a latent quantity during estimation. Only a limited number of contributions to the literature estimate uncertainty and its macroeconomic consequences jointly, and most are based on single country models. We analyze the special case of a shock restricted to the Euro area, where member states are highly related by construction. We find significant results of a decrease in real activity for all countries over a period of roughly a year following an uncertainty shock. Moreover, equity prices, short-term interest rates and exports tend to decline, while unemployment levels increase. Dynamic responses across countries differ slightly in magnitude and duration, with Ireland, Slovakia and Greece exhibiting different reactions for some macroeconomic fundamentals.
\end{abstract}

\bigskip
\begin{tabular}{p{0.2\hsize}p{0.65\hsize}} 
\textbf{Keywords:}  &Bayesian vector autoregressive models, factor stochastic volatility, uncertainty shocks\\
\end{tabular}

\smallskip
\begin{tabular}{p{0.2\hsize}p{0.4\hsize}}
\textbf{JEL Codes:} &C30, F41, E32 \\
\end{tabular}
\vspace{0.6cm}

\bigskip
\newpage

\section{Introduction}
There is consistent evidence in the literature that uncertainty levels increase after major economic and political events, affecting movements in key macroeconomic quantities. A specific example may be given by the Great Recession and the subsequent recovery, which was arguably slowed down by high levels of uncertainty \citep{leduc2016uncertainty}. In his seminal article, \citet{bloom2009impact} argues that economies are affected via depression of investment, consumption and employment, implying a fall in real activity and dampened productivity growth. Counteracting these unpleasant forces by traditional instruments like fiscal or monetary policy may prove difficult, as they lack effectiveness during periods of economic turmoil. In fact, such actions potentially induce further economic (policy) uncertainty. Consequently, decreasing the level of uncertainty may be considered a policy objective in its own right. A deep understanding of uncertainty and its dynamic effects on macroeconomic fundamentals appears to be crucial.

While some papers model uncertainty shocks in a dynamic stochastic general equilibrium and business cycle model context \citep[see, for instance,][]{fernandez2015fiscal,doi:10.3982/ECTA13960}, the bulk of the literature focuses on empirically assessing the impact of volatility shocks on the macroeconomy. Different measurements are employed, where a great number of articles relies on proxies. Popular approaches include the implied volatility of equity price returns, cross-sectional dispersion of firm profits, stock returns (for instance the volatility of S\&P 500 index options, VIX) or the occurrence of uncertainty-related keywords in leading newspapers \citep{bloom2009impact,dzielinski2012measuring,caggiano2014uncertainty,baker2016measuring,meinen2017measuring}. Moreover, uncertainty may be treated as latent quantity within the modeling framework, to simultaneously estimate arising implications for macroeconomic variables \citep{carriero2015impact,jurado2015measuring,crespo2017macroeconomic,mumtaz2016changing}. The framework we adopt in this paper follows the latter approach.

From an econometric point of view, \citet{bloom2009impact} captures uncertainty by using the implied volatility of equity price returns in a structural vector autoregressive model (VAR) and finds that an uncertainty shock generates a temporary drop in real activity (by the third month), rebound (usually by the sixth month) and long-run overshoot, which is different from the more persistent impact of first-moment shocks that generally take effect over at least a few quarters. A similar approach by \citet{carriero2015impact}, using a structural VAR including a measure of uncertainty as external instrument while simultaneously accounting for measurement errors, however, identifies the responses of macroeconomic quantities to an uncertainty shock to be more pronounced and persistent. 

Approaches to endogenously estimate uncertainty are mostly based on vector autoregressions introducing factors and stochastic volatility to model error terms. Intuitively, it is argued that the variable of interest is not financial market uncertainty, but real macroeconomic uncertainty.\footnote{That is, uncertainty observed in many economic time series at the same time across the private and public sector, corporations and households, and multiple economies. \citet{bloom2009impact}, for instance, identified almost 20 events inducing high uncertainty levels since the 1960s.} Findings obtained from studies in this spirit imply less, but longer periods of lowered real activity when compared to proxies based on financial market volatility \citep{jurado2015measuring}. Recent articles extend this basic framework to identify multiple types of uncertainty simultaneously, for instance, aggregate macroeconomic and financial uncertainty \citep{clark2016measuring}. 

Most studies consider a single country (typically the United States). Arguably, globalization and the synchronization of business-cycles have lead to the emergence of crucial interdependencies of economies \citep{stock2005understanding}. A real world example is again given by the Great Recession, with the crisis originating in the United States housing market quickly spreading to Europe and other parts of the industrialized world. It appears crucial to consider joint dynamics of multiple countries simultaneously to identify global transmission channels and avoid potential biases. \citet{crespo2017macroeconomic} employ a large-scale Bayesian VAR with stochastic volatility and novel shrinkage priors using data from G7 countries and find heterogeneous responses across countries.

This article contributes to the literature by identifying and estimating an uncertainty shock based on data from the Euro area and derives country-specific implications. We account for the emergence of synchronized clusters of economies and fill the blank space in the literature that is mostly concerned with the United States so far. For this purpose, we estimate a Bayesian VAR with factor stochastic volatility that allows us to capture uncertainty directly as latent quantity during estimation. The results are in line with the recent findings in this literature \citep{jurado2015measuring,crespo2017macroeconomic}. A significant decrease in real activity in most Euro area countries is observed over a period of roughly a year. Moreover, we find significant effects of uncertainty on unemployment and short-term interest rates, equity prices, as well as intra-European Union (EU) exports and exports to non-EU countries.

The remainder of this paper is organized as follows. \Cref{sec:methods} proposes the econometric framework employed to measure uncertainty in the Euro area. \Cref{sec:results} describes the data set and presents empirical results from an impulse response analysis and a forecast error variance decomposition. \Cref{sec:conclusion} concludes.

\section{Econometric framework}
\label{sec:methods}
\subsection{A Bayesian VAR model with factor stochastic volatility}
\label{sec:methods:1}

This section proposes the modeling framework. We follow \citet{crespo2017macroeconomic} and consider a standard VAR to estimate joint dynamics of macroeconomic and financial variables. The basic model is extended to include a factor structure with stochastic volatility in the error term, which allows us to estimate and identify a Euro area-specific shock. Specifically, the model is given by
\begin{equation}
 \label{eq:fullspec}
  \bm{y}_t = \sum_{p=1}^{P} \bm{B}_p \bm{y}_{t-p} + \bm{\epsilon}_t, \quad \bm{\epsilon}_t \sim \N(\bm{0},\bm{\Sigma}_t),
\end{equation}
where $\bm{y}_t$ is an $m$-dimensional vector including all observed time series across countries (that is, $m = cv$, where $c$ is the number of countries and $v$ the number of macroeconomic and financial variables). $\bm{B}_p (p=1,\hdots,P)$ are $m \times m$ dimensional matrices of regression coefficients with respect to the $p$-th lag of $\bm{y}_t$, and the error term $\bm{\epsilon}_t$ is assumed to follow a Gaussian distribution with zero mean and time-varying variance covariance matrix $\bm{\Sigma}_t$. 

Following \citet{aguilar2000bayesian}, we decompose the error term into a structure featuring common factors that capture shared movements in volatility plus idiosyncratic shocks,
\begin{equation}
  \label{eq:errorspec}
  \bm{\epsilon}_t = \bm{X}\bm{f}_t + \bm{\eta}_t,
\end{equation}
with $\bm{X}$ being an $m \times q$ matrix of factor loadings, while $\bm{f}_t \sim \N(\bm{0},\bm{H}_t)$ is a vector containing $q$ latent common factors. The time-varying variance-covariance matrix of the factors $\bm{H}_t = \text{diag}(h_{1t},\hdots,h_{qt})$ is a diagonal matrix. The respective elements $h_{it}$ ($i=1,\hdots,q$) are assumed to follow univariate stochastic volatility processes, that is, the log-volatilities follow a centered AR(1) process,
\begin{equation}
\label{eq:stochvolspec}
  \log(h_{it}) = \mu_i^{(h)} + \phi_i^{(h)} (\log(h_{it-1}) - \mu_i^{(h)}) + \xi_{it}^{(h)}. 
\end{equation}
Hereby, $\mu_i^{(h)}$ denotes the mean of the log volatilities, $\phi_i^{(h)} \in (-1,1)$ is the persistence parameter and $\xi_{it}^{(h)} \sim \N(0,\Xi_i^{(h)})$. Moreover, $\bm{\eta}_t \sim \N(\bm{0}, \bm{\Omega}_t)$ is a zero mean Gaussian error term, where we again impose a time-varying variance-covariance matrix $\bm{\Omega}_t = \text{diag}(\omega_{1t},\hdots,\omega_{mt})$. The elements $\omega_{jt}$ ($j=1,\hdots,m$) follow a centered AR(1) process analogous to \autoref{eq:stochvolspec},
    \begin{equation}
          \log(\omega_{jt}) = \mu_j^{(\omega)} + \phi_j^{(\omega)} (\log(\omega_{jt-1}) - \mu_j^{(\omega)}) + \xi_{jt}^{(\omega)}.
    \end{equation}
The variance-covariance matrix of $\bm{\epsilon}_t$ is thus given by $\bm{\Sigma}_t = \bm{X} \bm{H}_t \bm{X}' + \bm{\Omega}_t$, where time-variation in $\bm{\Sigma}_t$ stems both from the stochastic volatility specification of the factors $\bm{f}_t$ and the idiosyncratic error variances in $\bm{\Omega}_t$. Standard information criteria suggest it is sufficient to identify one common factor in the error term. This implies that $\bm{f}_t$ and $\bm{H}_t$ are scalars, and the factor loadings matrix reduces to a column vector. Econometric identification of the model is achieved by setting the first element of the vector $\bm{X}$ to be equal to one, based on \citet{aguilar2000bayesian}.

\subsection{Prior specification}
\label{sec:methods:2}
The proposed model is highly parameterized and requires Bayesian methods to be estimated. This entails elicitation of suitable prior distributions on the parameters. For the VAR coefficients, we employ the Normal-Gamma shrinkage prior popularized by \citet{griffin2010} and extended to a VAR context by \citet{huber_feldkircher}. It is useful to stack the elements of the coefficient matrices to obtain $\bm{\beta}_p = \text{vec}(\bm{B}_p)$, where in the following $\beta_{ip}$ denotes the $i$-th element of the $k$-dimensional vector $\bm{\beta}_p$, where we define $k = m^2$. The employed hierarchical approach termed global-local shrinkage prior introduces a set of idiosyncratic local scaling parameters $\tau_{ip}$ and lag-specific shrinkage parameters $\lambda_p^2$. A standard variant is given by a scale mixture of a zero-mean Gaussian distribution with Gamma mixing density,
\begin{equation}
\beta_{ip} | \tau_{ip} \sim \N \left(0, 2\tau_{ip}/\lambda_p^2\right), \quad \tau_{ip} \sim G(\kappa_p,\kappa_p).
\end{equation}
For the lag specific shrinkage parameters $\lambda_p^2$, we follow \citet{huber_feldkircher} and assume that shrinkage increases with lag order.\footnote{This idea is based on the well-known Minnesota prior \citep{literman1986,sims}. However, instead of pushing the system towards a random walk specification, we shrink all coefficients towards zero to guarantee the stability of our model.} Consequently, we assume $\lambda_p^2$ to be the product of independent Gamma priors, 
\begin{equation}
  \lambda_p^2 = \prod_{j=1}^p \delta_j, \quad \delta_n \sim G(c_j,d_j),
\end{equation}
where $c_j$ and $d_j$ ($j = 1,\hdots,p$) are hyperparameters. If each $\delta_j$ exceeds unity, the prior variance decreases, implying that coefficient matrices of higher lag orders are pushed more strongly towards sparsity. Regarding the factor loadings, we follow the standard approach and use independent zero-mean Gaussian priors with weakly informative variance. Finally, for the prior specification of the stochastic volatility processes in our model, we follow \citet{schnatterkastner2014}.\footnote{Estimation is carried out using the \textit{R}-package provided by \citet{kastner2016dealing}.} An overview on prior hyperparameter values is given in \cref{app:hyperparameters}.

\subsection{Posterior simulation}
\label{sec:methods:4}
We now turn to a description of the Markov chain Monte Carlo algorithm we employ. Rewriting the model from \autoref{eq:fullspec} allows for equation-by-equation estimation, exploiting the diagonal structure of $\bm{\Omega}_t$. Conditional on $\bm{X}\bm{f}_t$, $\bm{\hat{y}}_t = \bm{y}_t - \bm{X}\bm{f}_t$ results in
\begin{equation}
    \bm{\hat{y}}_t = \sum_{p=1}^{P} \bm{B}_p \bm{y}_{t-p} + \bm{\eta}_t,
\end{equation}
which is a system of unrelated regressions. Simulating most related quantities is standard in the Bayesian literature and hence not derived and discussed in detail. For the sake of completeness, we briefly sketch the Gibbs sampler that is employed. Conditional on all other parameters of the model, the VAR coefficients are sampled. Subsequently, we draw $\bm{\Omega}_t$ conditional on the latent factors $\bm{f}_t$, the factor loadings matrix $\bm{X}$ and the full history of log-volatilities. The latent factors are then simulated from independent normal distributions \citep[for further reference, see][]{aguilar2000bayesian}. Afterwards, we sample the factor loadings. Estimating the shrinkage parameters $\tau_{ip}$ and $\lambda_p^2$ for the Normal-Gamma shrinkage prior is discussed in extensive detail in \citet{huber_feldkircher}. The local parameter $\tau_{ip}$ is sampled from a generalized inverse Gaussian (GIG) distribution,
    \begin{equation}
      \tau_{ip} | \bullet \sim \text{GIG}(\kappa_p-1/2,\beta_{ip}^2,\kappa_p\lambda_p^2).
    \end{equation}
The global shrinkage parameter $\lambda_p^2$ is sampled from a Gamma distribution with
\begin{equation}
  \lambda_p^2 | \bullet \sim 
  \begin{cases}
  G\left(c_1 + \kappa_1 k, d_1 + \kappa_1 \sum_{i=1}^{k} \tau_{i1}/2\right) \quad &\text{for } p=1\\
  G\left(c_p + \kappa_p k, d_p + \kappa_p (\prod_{z=1}^{p-1}\lambda_{z}) (\sum_{i=1}^{k} \tau_{ip}) /2\right) \quad &\text{for } p>1.
  \end{cases}
\end{equation}
Additional details on the derivations of these conditional posterior distributions may be found in \citet{huber_feldkircher} and provided supplementary material.

\section{Empirical results}
\label{sec:results}
In this section we briefly describe our data set and present the obtained common uncertainty factor. Moreover, to assess the quantitative impact of uncertainty shocks, we discuss the share of explained forecast error variances. The section is completed by an impulse response analysis of the variables in our system, where we present the reaction of various macroeconomic quantities to a common uncertainty shock in the Euro area.

\subsection{Data}
The data set is comprised of 12 out of 19 Euro Area countries, namely Austria, Belgium, Germany, Greece, Spain, Finland, France, Ireland, Italy, the Netherlands, Portugal and Slovakia. We exclude the Baltic states, Malta, Cyprus, Slovenia and Luxembourg due to the lack of appropriate time series. In addition, the data set contains aggregated quantities for the whole Euro area (EA19) to capture overall trends. The final time series covers the period from January 2000 to September 2015 on a monthly basis. We include standard macroeconomic quantities for all countries, that is, gross domestic product (GDP), consumer prices, equity prices, unemployment and interest rates, and exports.  Moreover, oil prices are introduced as an exogenous variable to control for structural changes in the global economy. Data sources include Eurostat, the European Commission, the Organisation for Economic Co-operation and Development and the Oesterreichische Nationalbank. A comprehensive description of the data and corresponding sources can be found in \cref{app:data-app}.

\subsection{Measures of uncertainty}
\begin{figure}[!htbp]     
	\includegraphics[width=1\textwidth]{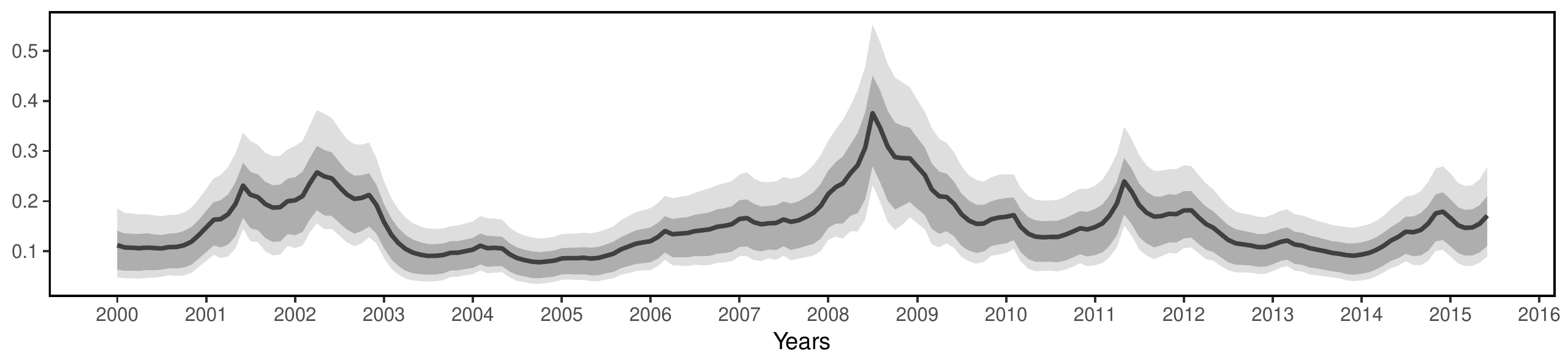}
	\caption{Volatility of posterior median of the latent factor.}
  \caption*{\footnotesize{\textit{Note}: The obtained measure is depicted by the dark grey line, indicated confidence bands are the 16$^{\text{th}}$ and 84$^{\text{th}}$ (grey area), and 5$^{\text{th}}$ and 95$^{\text{th}}$ percentiles (light grey area), respectively.}}
	\label{uncertainty_common-factor}
\end{figure}

An overview of volatility in Euro area economies is given in \autoref{uncertainty_common-factor}, which depicts the variance of the uncertainty factor over time. The millennium begins with the burst of the US based dot-com bubble reaching Europe. Roughly at the same time, the 9/11 terror attack adversely impacts the economic environment and financial markets (causing the Dow Jones to fall 1.370 points, a loss of \$1.4 trillion in market value). This period of comparatively high volatility is followed by a few years of relative tranquility. The global financial crisis hitting Europe is evidenced by the peak in early 2008, followed by the European Debt Crisis two years later.

\begin{figure}[!htbp]     
\centering
\begin{subfigure}{\linewidth}
  \centering
  \includegraphics[width=1\textwidth]{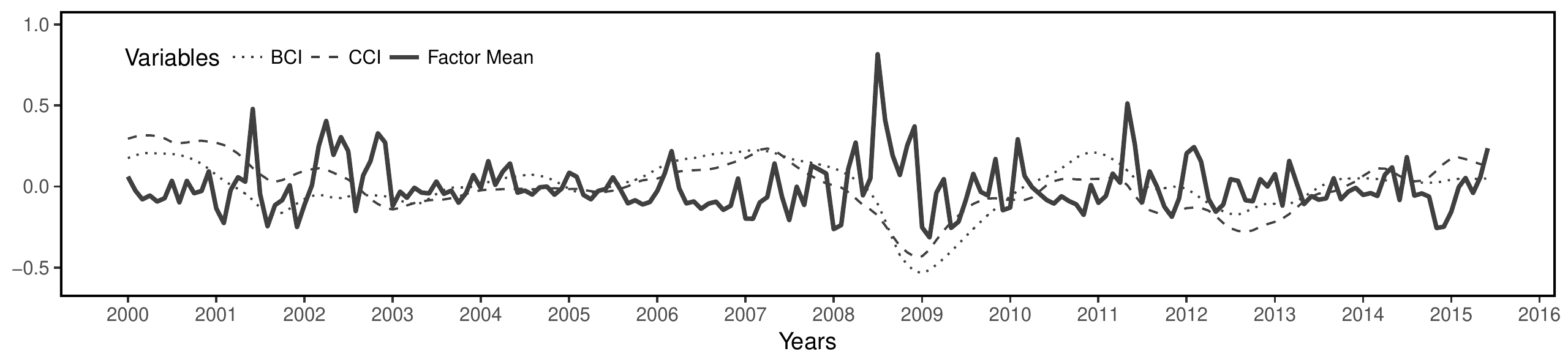}
  \caption{Consumer and business confidence index}
  \label{uncertainty_measures_cci-bci} 
\end{subfigure}
\begin{subfigure}{\linewidth}
  \centering
  \includegraphics[width=1\textwidth]{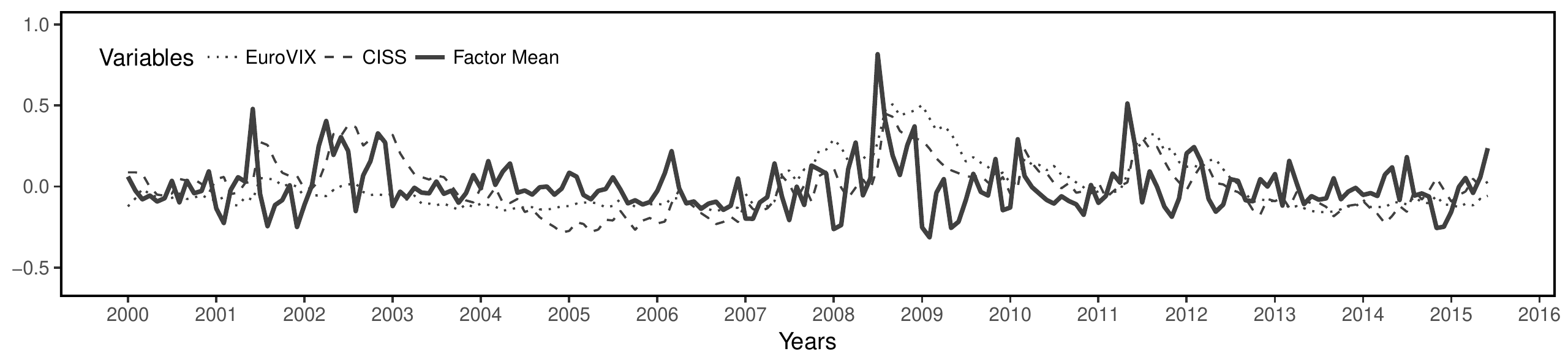}
  \caption{Proxies for uncertainty}
  \label{uncertainty_measures_eurovix-ciss} 
\end{subfigure}
\caption{Comparison of the estimated uncertainty factor and other measures.}
\label{uncertainty_measures}
\end{figure}

To assess how well we this measure coincides with related quantities, we compare the estimated latent factor to commonly known proxies of (mostly financial market) volatility and economic stability. Confidence in the economy of both consumers and business owners is typically expected to decrease in times of high uncertainty. Indeed, as shown in \autoref{uncertainty_measures_cci-bci}, we find that both the Consumer Confidence Index (CCI) and Business Confidence Index (BCI) published by the European Commission show a lagged countercyclical behavior during peak-levels of uncertainty. This is a strong indicator that uncertainty has real, measurable impacts on the economy.

\Cref{uncertainty_measures_eurovix-ciss} compares our measure of volatility with two further common proxies. On the one hand, the EuroVIX, which is the volatility index based on the EUROSTOXX stock markets, on the other hand, the Composite Indicator of Systemic Stress (CISS), an index constructed by the European Central Bank (ECB) to capture the exposure of the European system to economic fluctuations. Again, the estimated latent factor tracks the estimated proxies quite well. The two chosen indices mostly reflect the situation on financial markets, which may imply that our measure of uncertainty is tied to financial market volatility. However, the estimated quantity deviates from the other measures in the years before the global financial crisis. This suggests that our uncertainty measure is in fact able to capture more than just financial uncertainty.

\subsection{Explaining innovation variance by the uncertainty factor}
\begin{figure}[!htbp]
\centering
\begin{subfigure}{\linewidth}
  \centering
  \caption{Real gross domestic product (GDP)}
  \includegraphics[width=1\textwidth]{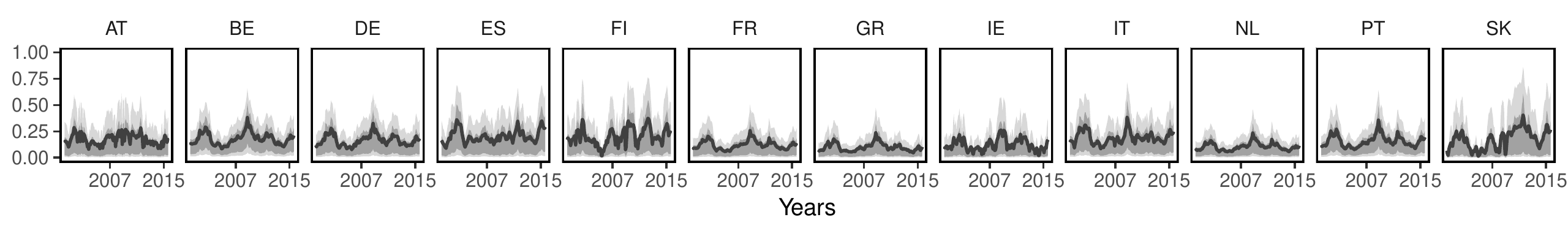}
  \label{shares_y} 
\end{subfigure}
\begin{subfigure}{\linewidth}
  \centering
  \caption{Unemployment}
  \includegraphics[width=1\textwidth]{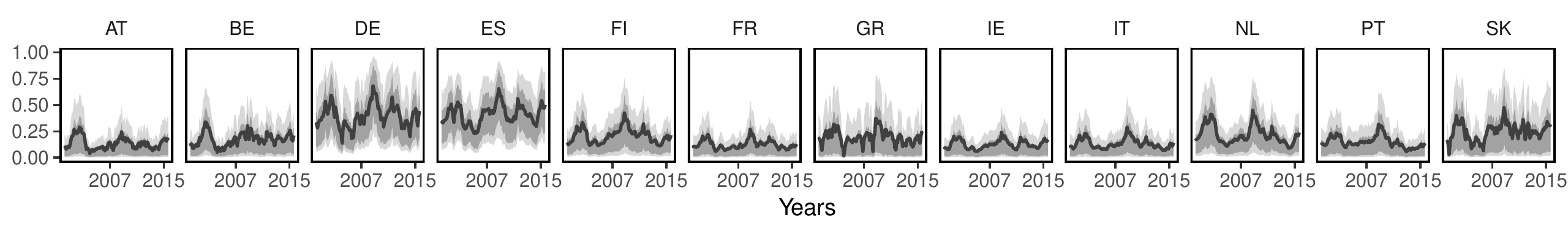}
  \label{shares_unempr} 
\end{subfigure}
\begin{subfigure}{\linewidth}
  \centering
  \caption{Short-term interest rates}
  \includegraphics[width=1\textwidth]{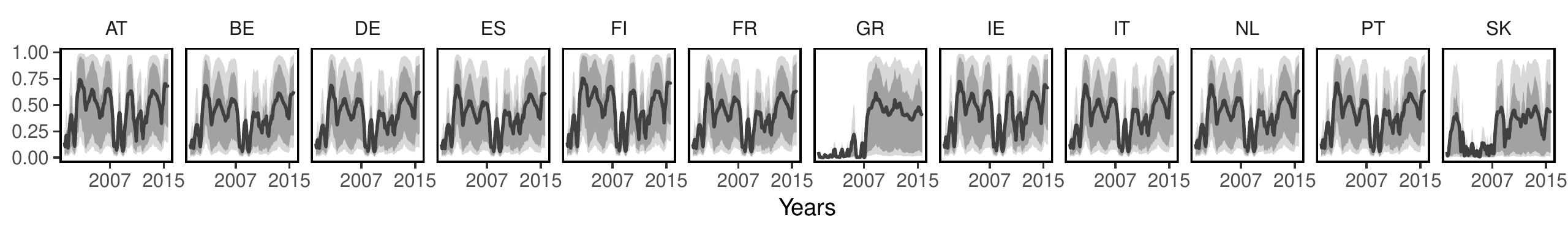}
  \label{shares_stir} 
\end{subfigure}
\begin{subfigure}{\linewidth}
  \centering
  \caption{Consumer prices}
  \includegraphics[width=1\textwidth]{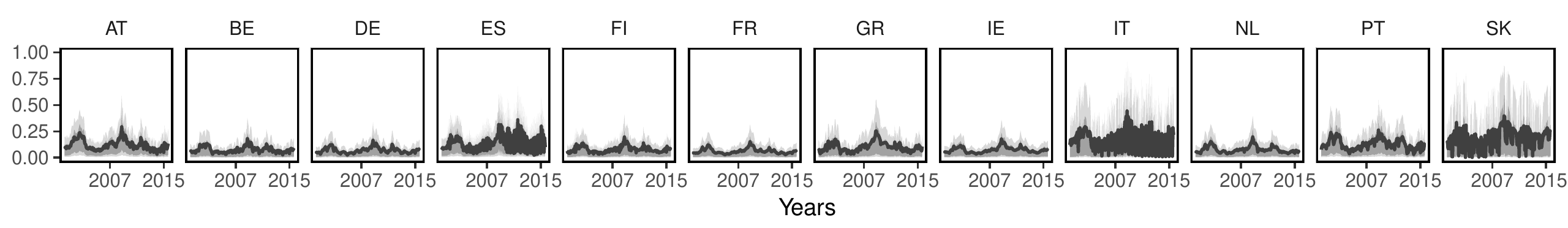}
  \label{shares_p} 
\end{subfigure}
\begin{subfigure}{\linewidth}
  \centering
  \caption{Equity prices}
  \includegraphics[width=1\textwidth]{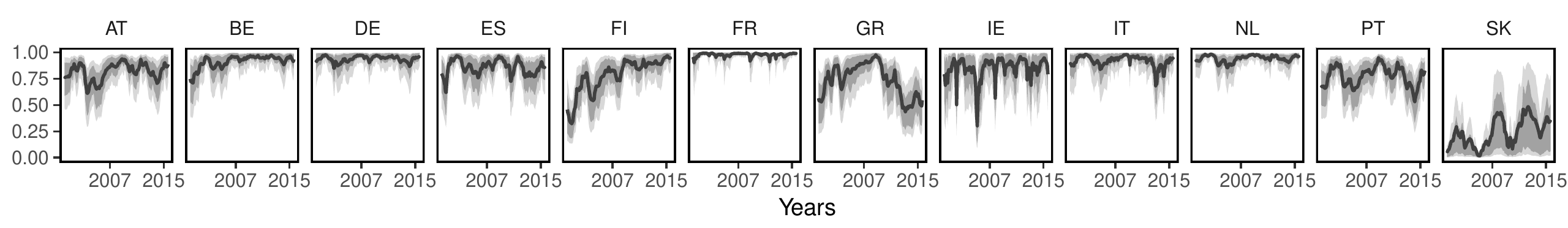}
  \label{shares_eq} 
\end{subfigure}
\begin{subfigure}{\linewidth}
  \centering
  \caption{Intra-EU exports}
  \includegraphics[width=1\textwidth]{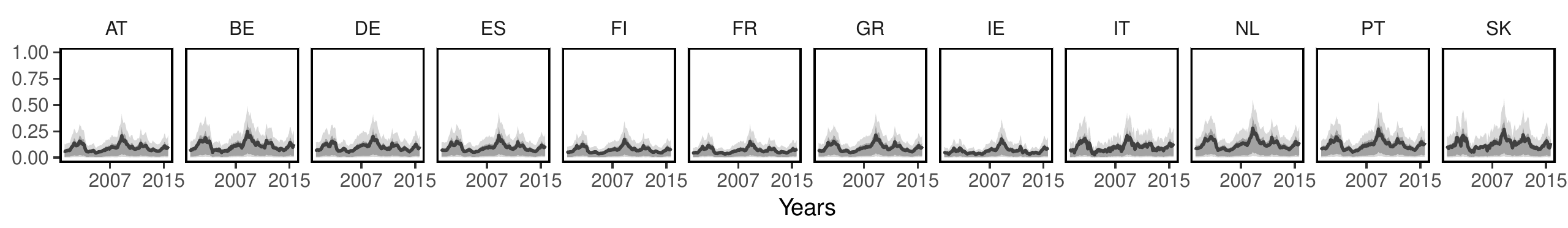}
  \label{shares_exintra} 
\end{subfigure}
\caption{Share of innovation variance explained by uncertainty factor}
\caption*{\footnotesize{\textit{Note}: The obtained measure is depicted by the dark grey line, indicated confidence bands are the 16$^{\text{th}}$ and 84$^{\text{th}}$ (grey area), and 5$^{\text{th}}$ and 95$^{\text{th}}$ percentiles (light grey area), respectively. Country codes may be found in \cref{app:data-app}.}}
\label{shares}
\end{figure}

To gauge the quantitative importance of uncertainty shocks regarding overall macroeconomic dynamics, we investigate how much of the forecast error variance of the variables in the system can be explained by the latent factor. \Cref{shares} displays proportions of the explained innovation variance over time for all variables except for non-EU exports.\footnote{We refrain from presenting plots of this variable due to the similarity with intra-EU exports and space limitations.}

\Cref{shares_y} shows the forecast error variance decomposition for real GDP, where between 15 and 30 percent of the innovation variance is explained by the latent factor. An interesting pattern arises, as the variance due to the uncertainty factor increases in times of economic turmoil. This phenomenon occurs across all countries, with the dot-com bubble and Great Recession clearly visible. However, it must be acknowledged that the European sovereign debt crisis is not observable to a great extent. \Cref{shares_unempr} depicts the explained forecast error variance of unemployment. Again, on average between 15 and 25 percent of the innovation variance can be explained by the latent factor. Germany and Spain represent two outliers with more variation in explained shares. A plausible explanation might be a more severe reaction of the labor market within these countries due to financial uncertainty. The share of explained variance for short-term interest rates in \autoref{shares_stir} shows fierce movements, varying from 10 to 75 percent, with high values during times of economic prosperity and decreases during economic downturns. \Cref{shares_p} presents the comparatively low shares (below 15 percent for most countries) of explained variance for consumer prices. We conjecture that price forecast variance is mostly driven by other factors.

Concerning equity prices, the latent factor explains almost all of the forecast error variance as shown in \autoref{shares_eq}. The only exceptions are Greece and Slovakia. For the latter, the specific behavior may be due to the lower dependence of Central and Eastern European countries on the global financial market, while Greece arguably takes a special place considering the aftermath of the Great Recession. For intra-EU exports (and EU-outbound exports) in \autoref{shares_exintra}, we observe a comparatively low level of explained forecast error variance. Identifying a general pattern, we find that the explained fractions typically increase for most variables in periods of financial market turmoil.

\subsection{Impulse responses of uncertainty shocks}
Impulse response functions are calculated by simulating a shock to the common uncertainty factor. A shock, or any change in $\bm{f}_t$ is transmitted to the other variables through the respective factor loadings. Fast moving variables typically exhibit higher loadings, as they are expected to react to a shock immediately. Following the approach by \citet{crespo2017macroeconomic}, we scale the shock to represent a ten percent decline of equity prices on average. Scales of the responses may be interpreted as follows. For variables in logarithms, the scale represents deviations from the level at the time of the shock in percent. This is the case for all quantities except short-term interest rates and the unemployment rate. The scale regarding these variables constitutes a deviation from the rate at the time of the shock in percentage points.

\begin{figure}[!htbp]
	\includegraphics[width=1\textwidth]{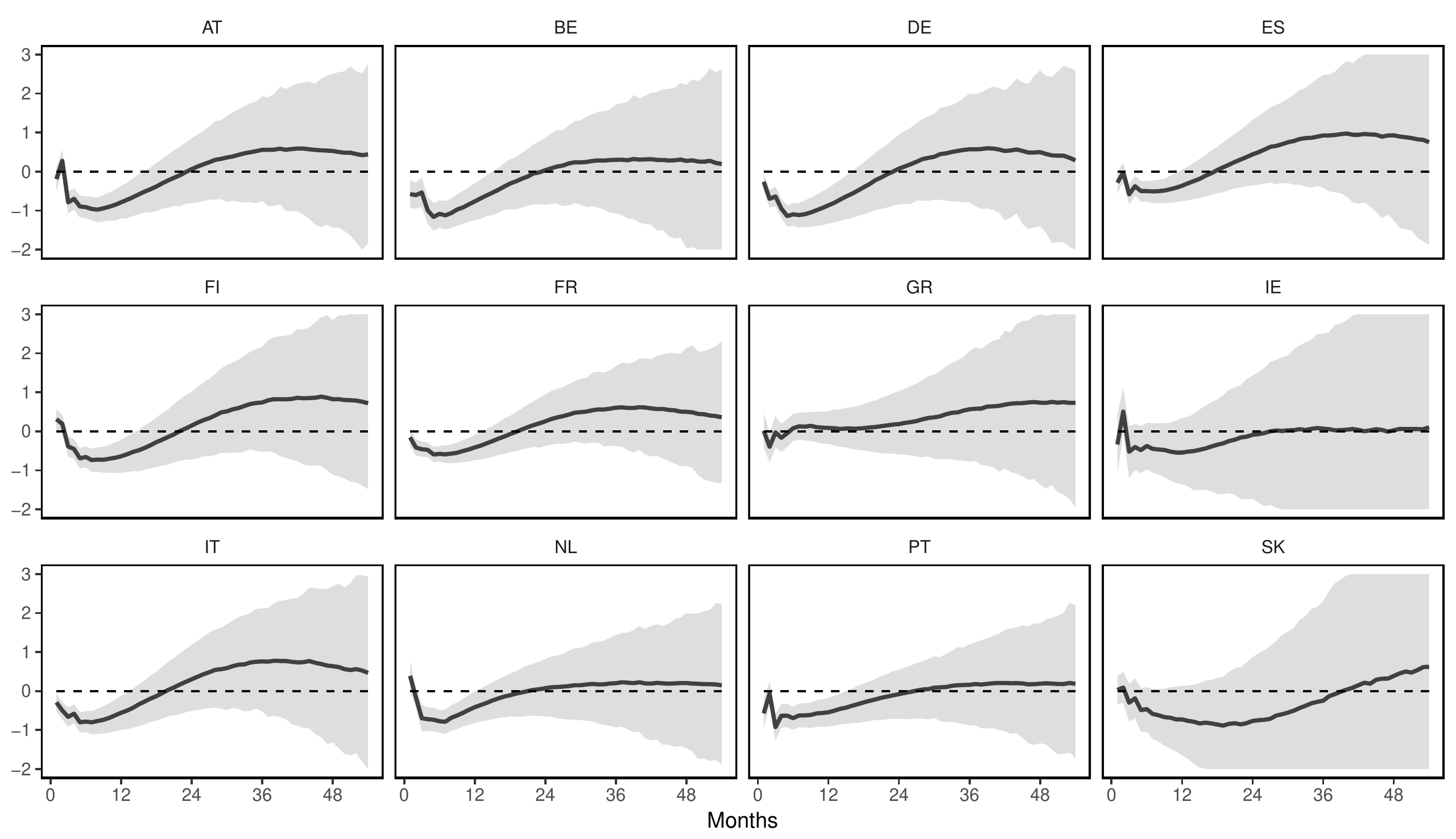}
	\caption{Impulse response functions of GDP as measure for real activity in Euro area countries.}
  \caption*{\footnotesize{\textit{Note}: Posterior distribution of impulse responses in percent. The median is depicted in grey, the 16$^{\text{th}}$ and 84$^{\text{th}}$ percentile are indicated by the light grey area. The dashed line indicates zero. Country codes may be found in \cref{app:data-app}.}}
	\label{irf_factor_y}
\end{figure}

The observed behavior of real activity in terms of GDP in \autoref{irf_factor_y} is consistent with recent findings in both the theoretical and empirical literature. For some countries, we find significant decreases on impact. The period of depressed real activity lasts for roughly 10 to 12 months; from that point onward responses turn insignificant. This is a major difference to the empirical findings of \citet{bloom2009impact}, who presents evidence for a statistically significant medium-run real activity overshoot. Hence, our empirical results regarding GDP provide further evidence for the absence of a significant rebound effect. A puzzle emerges for Greece, which shows no significant results over the impulse response horizon considered.

\begin{figure}[!htbp]
	\includegraphics[width=1\textwidth]{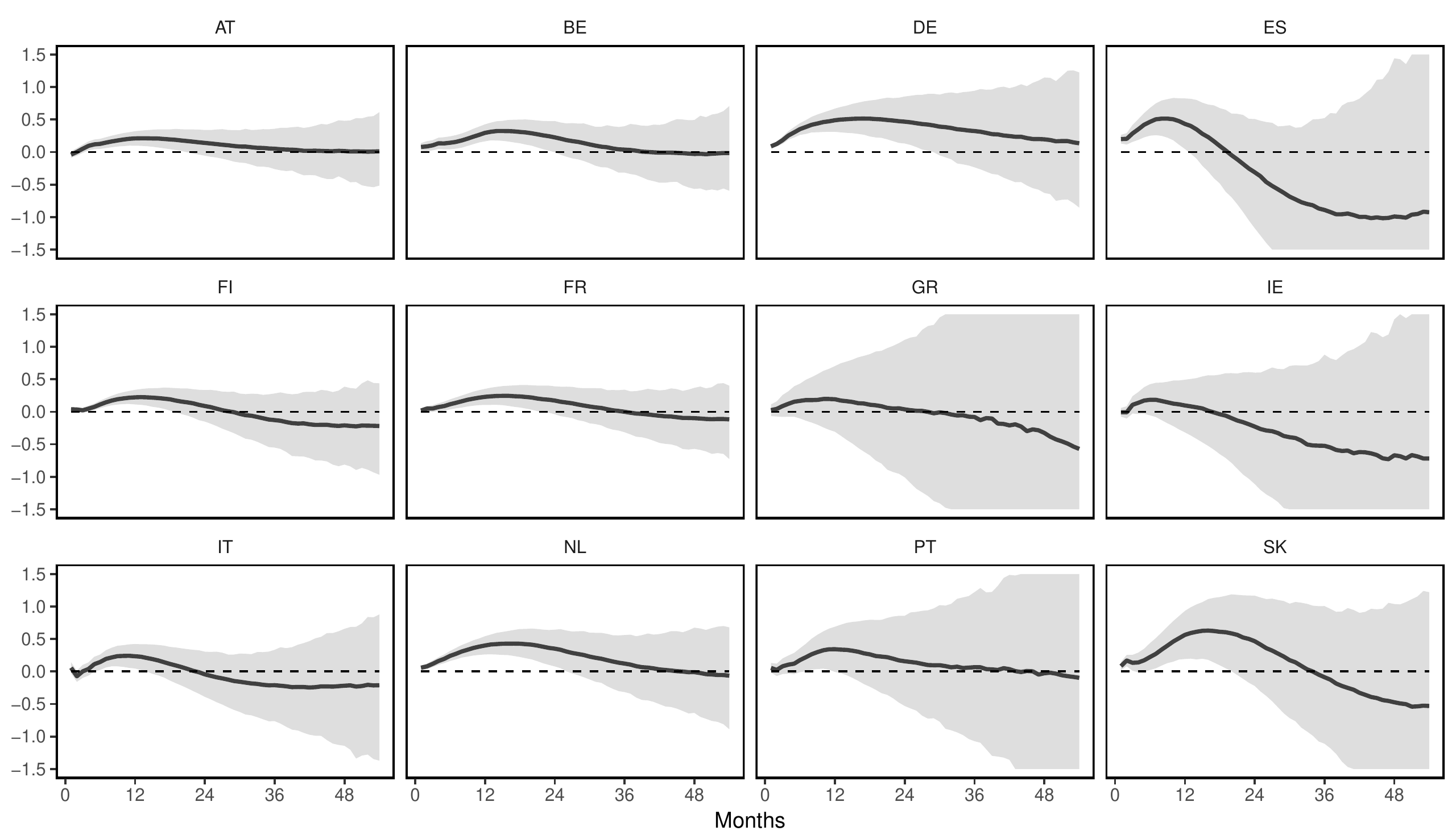} 
	\caption{Impulse response functions of the unemployment rate in Euro area countries.}
  \caption*{\footnotesize{\textit{Note}: Posterior distribution of impulse responses in percentage points. The median is depicted in grey, the 16$^{\text{th}}$ and 84$^{\text{th}}$ percentile are indicated by the light grey area. The dashed line indicates zero. Country codes may be found in \cref{app:data-app}.}}
	\label{irf_factor_unempr}
\end{figure}

Movements in the unemployment rate (depicted in \autoref{irf_factor_unempr}) corroborate the theoretical channel and follow the reasoning with respect to real activity. Uncertainty causes firms to stop hiring new employees, hence there is virtually no effect on impact. Over the course of time people are laid off, corresponding to profit maximization considerations of corporations. Compared to already established research results from \citet{jurado2015measuring}, we find similar responses, but the effects we observe are less persistent. Increased unemployment occurs between approximately 12 and 24 months after impact of the uncertainty shock. Effects tractable for comparatively longer periods are observed in Germany, Slovakia and the Netherlands, but also in the cases of Spain and Portugal. A less pronounced, but still significant increase of unemployment occurs in Austria, Belgium, Finland and France. The results from Greece, Italy and Ireland are not significant over the horizon considered and much broader confidence bounds were estimated for these three countries.

\begin{figure}[!htbp]
	\includegraphics[width=1\textwidth]{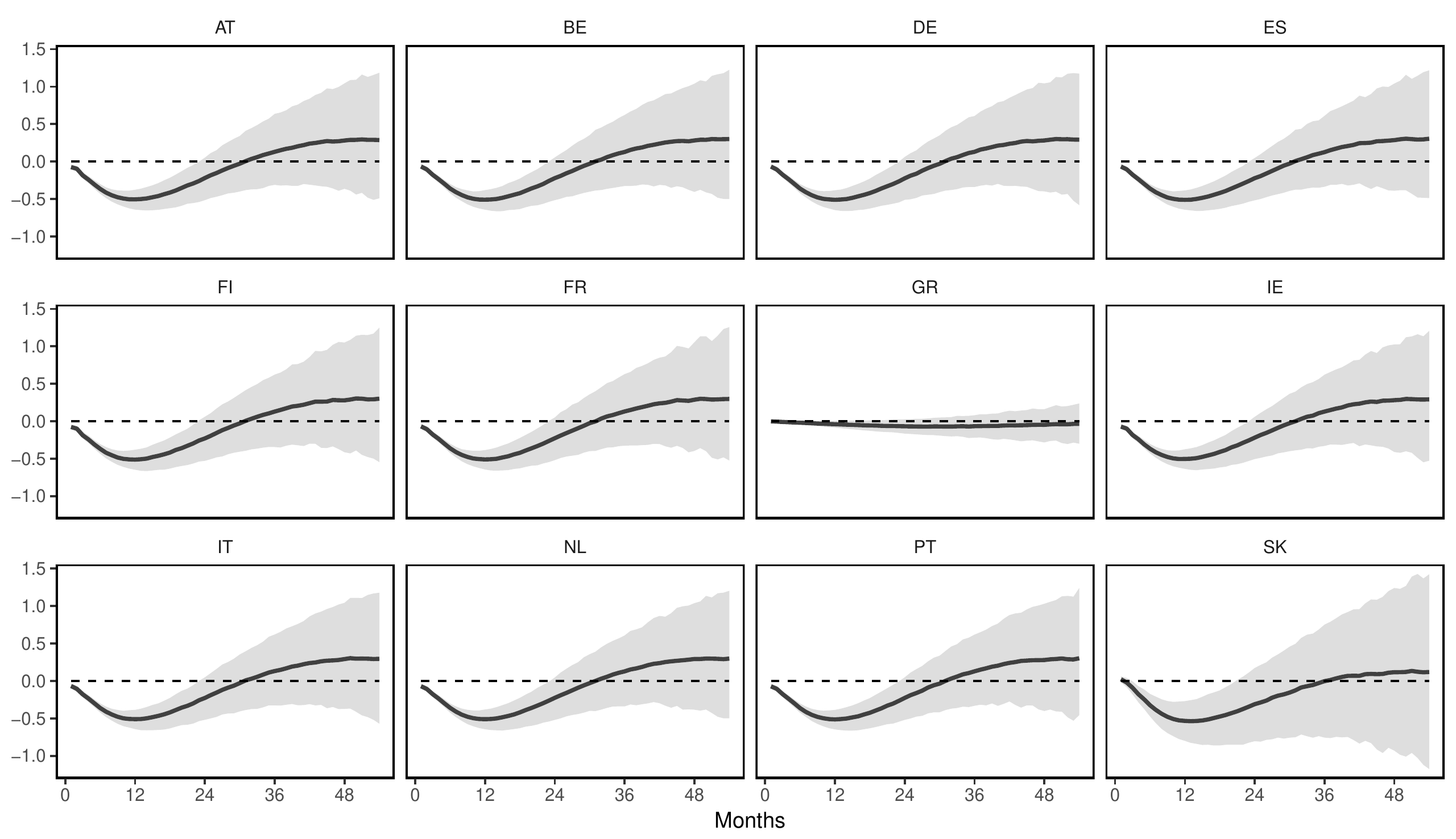}
	\caption{Impulse response functions of the short-term interest rate in Euro area countries.}
  \caption*{\footnotesize{\textit{Note}: Posterior distribution of impulse responses in percentage points. The median is depicted in grey, the 16$^{\text{th}}$ and 84$^{\text{th}}$ percentile are indicated by the light grey area. The dashed line indicates zero. Country codes may be found in \cref{app:data-app}.}}
	\label{irf_factor_stir}
\end{figure}

Impulse responses of the short-term interest rate are depicted in \autoref{irf_factor_stir}. Generally, the dynamic response is homogeneous across most countries considered, except Greece. The shape, magnitude and duration of the lowered interest rates, reaching their peak at -$0.5$ percentage points after roughly 12 months is shared by most countries we consider. After two years, the short-term interest rate change based on the uncertainty shock turns insignificant. Comparing these results to previous studies, this response is more persistent than found in the literature. It appears reasonable that central banks respond to an uncertainty shock by lowering interest rates to foster investment and restore economic prosperity and growth. As all countries depend on the ECB, similar reactions are to be expected. However, the case of Greece appears to require further attention. A plausible explanation is that standard transmission channels of monetary policy actions were impeded. 

\begin{figure}[!htbp]
	\includegraphics[width=1\textwidth]{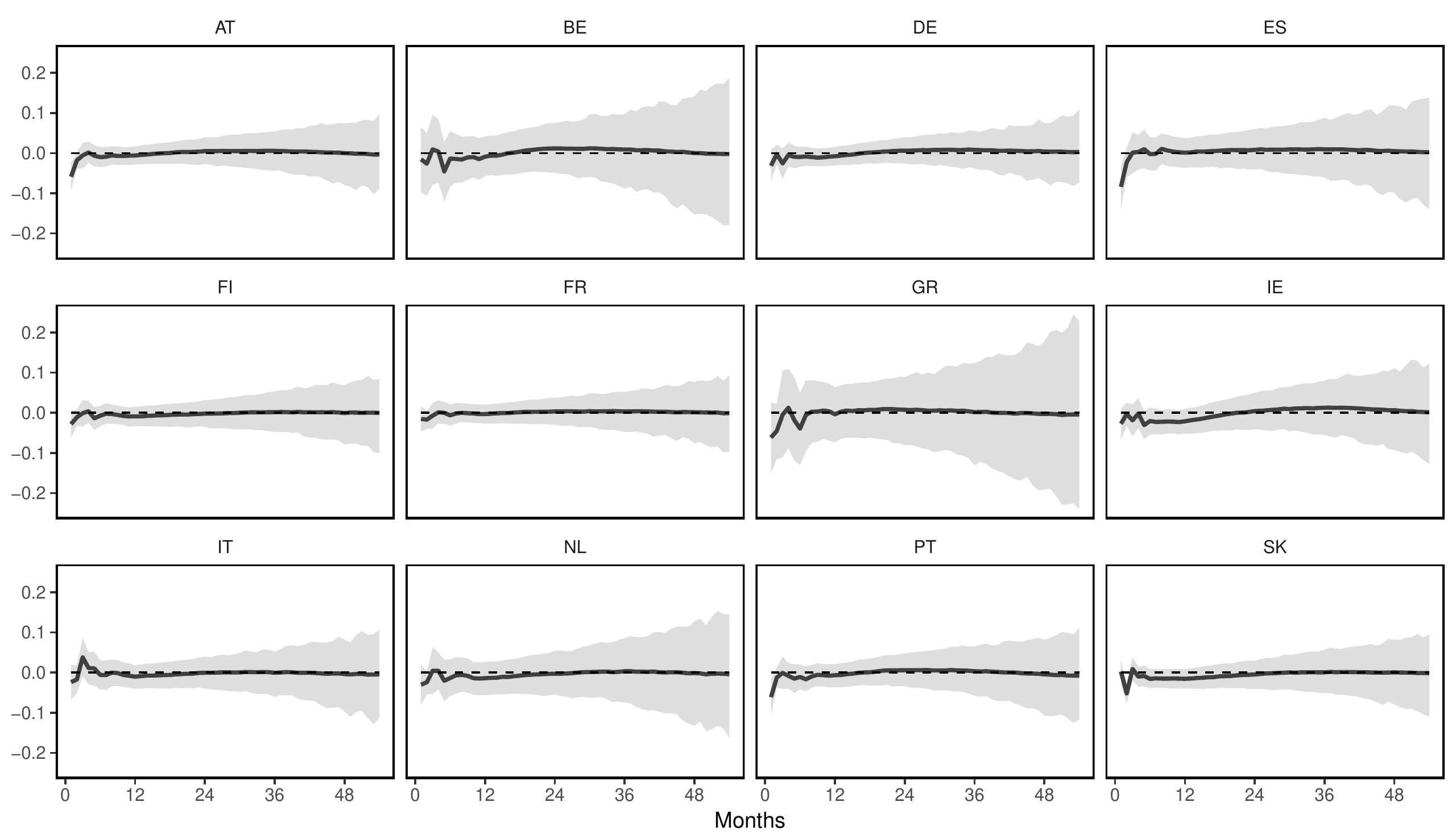}
	\caption{Impulse response functions of price level in Euro area countries.}
  \caption*{\footnotesize{\textit{Note}: Posterior distribution of impulse responses in percent. The median is depicted in grey, the 16$^{\text{th}}$ and 84$^{\text{th}}$ percentile are indicated by the light grey area. The dashed line denotes zero. Country codes may be found in \cref{app:data-app}.}}
	\label{irf_factor_p}
\end{figure}

No significant reactions of consumer prices to an uncertainty shock occur (see \autoref{irf_factor_p}). Main theoretical channels explaining this behavior are based on the notion of counteracting forces. \citet{fernandez2011risk} identify the transmission to be driven by the aggregate demand channel, where uncertainty implies reduced consumption demand of households, thereby decreasing prices. However, via the upward pricing bias channel, the overall price level might be increased due to profit-maximizing firm behavior.

\begin{figure}[!htbp]
	\includegraphics[width=1\textwidth]{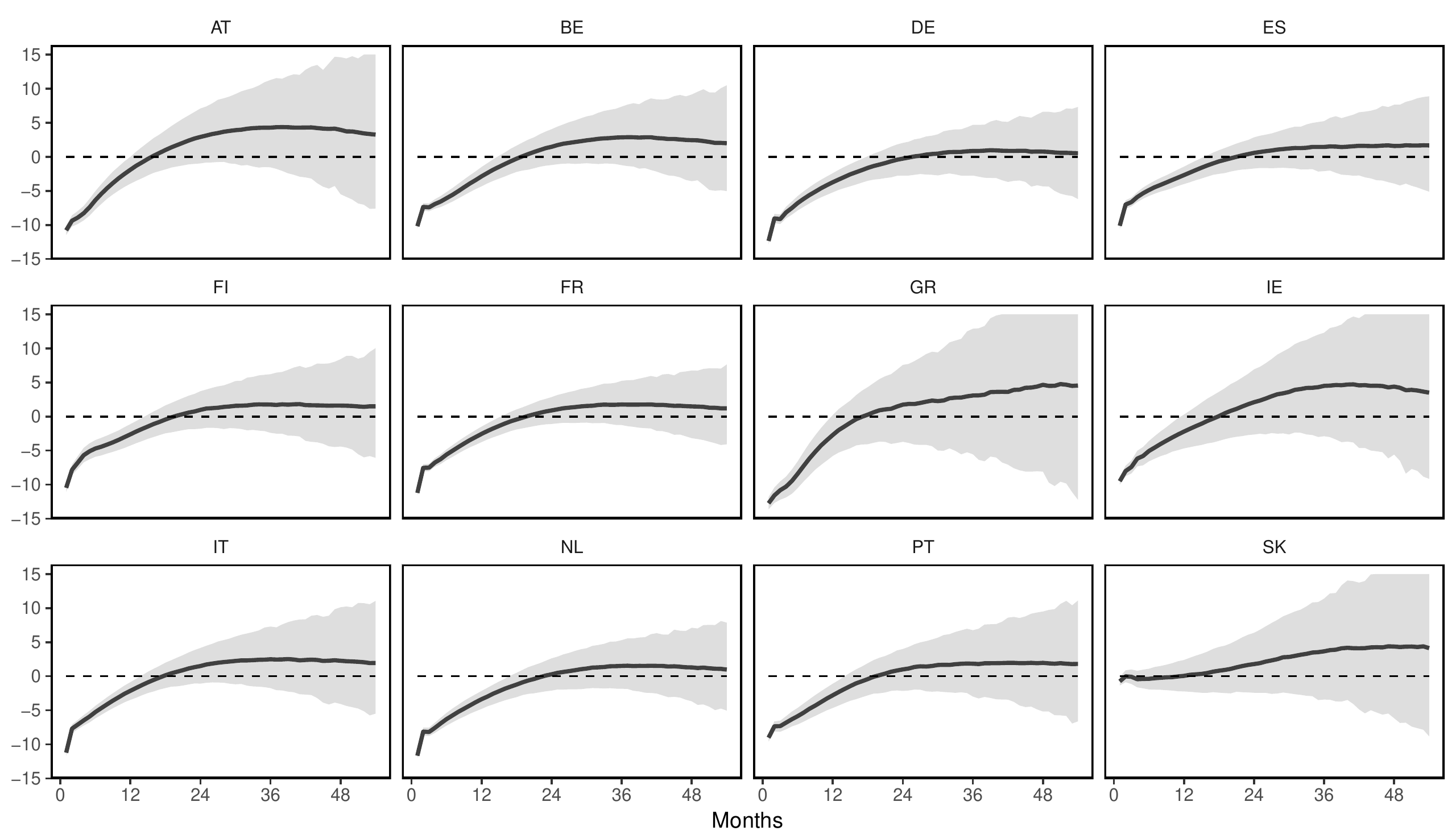}
	\caption{Impulse response functions of equity prices in Euro area countries.}
  \caption*{\footnotesize{\textit{Note}: Posterior distribution of impulse responses in percent. The median is depicted in grey, the 16$^{\text{th}}$ and 84$^{\text{th}}$ percentile are indicated by the light grey area. The dashed line indicates zero. Country codes may be found in \cref{app:data-app}.}}
	\label{irf_factor_eq}
\end{figure}

Impulse responses for equity prices are depicted in \autoref{irf_factor_eq}. Apart from Slovakia, they react with a sharp decrease to an uncertainty shock in all countries. By construction, the average drop across countries on impact corresponds to ten percent. Responses are significantly different from zero for at least 12 months in most countries, and up to 18 months in some cases. A reason explaining the high impact of uncertainty on equity prices is given by the variables fast moving nature. That is, equity prices react faster and to a bigger extent than other, slower moving variables like unemployment. The result regarding equity prices corroborate findings from the literature \citep{clark2016measuring,crespo2017macroeconomic}.

\begin{figure}[!htbp]
	\includegraphics[width=1\textwidth]{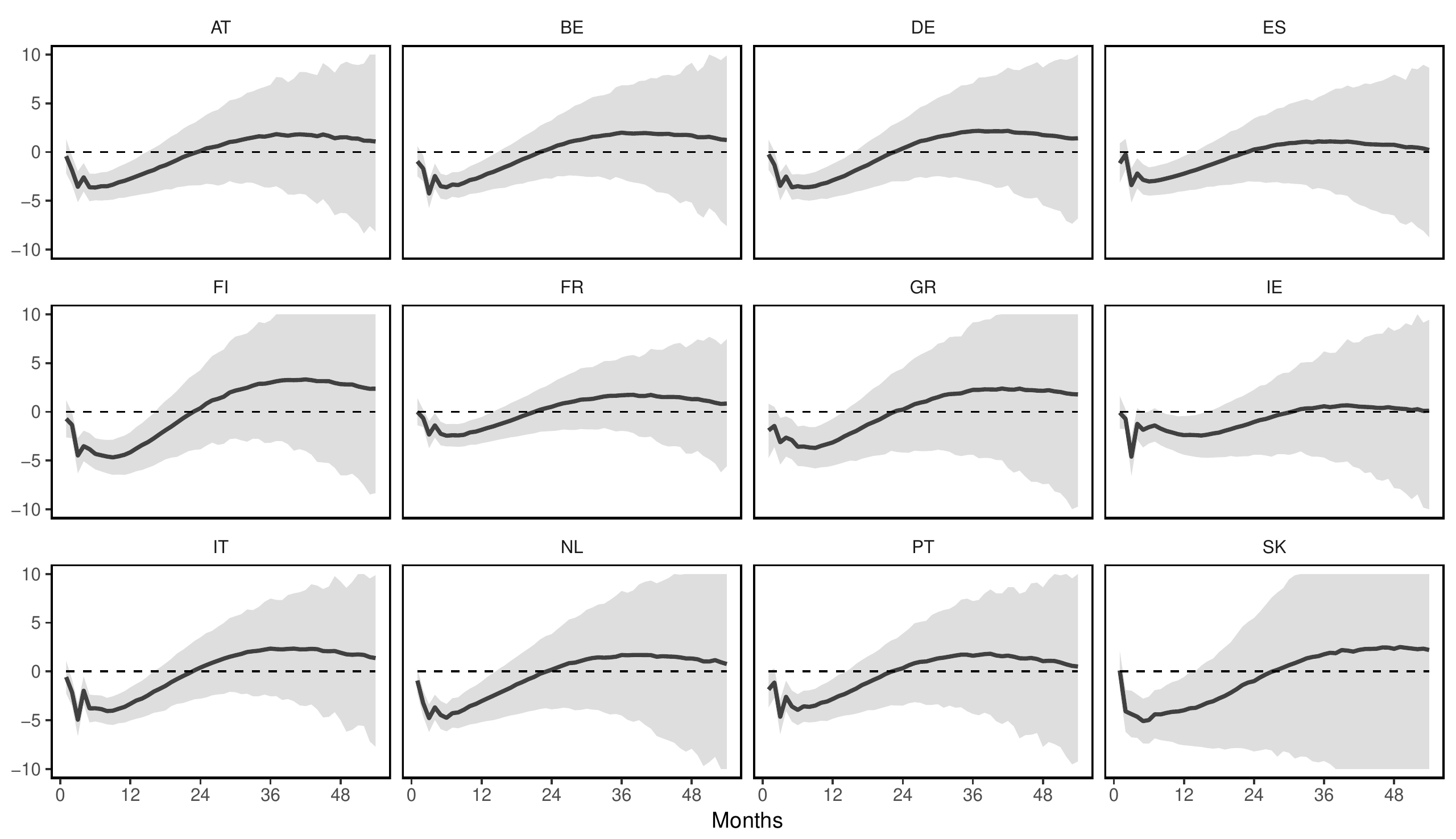}
	\caption{Impulse response functions of exports among EU countries.}
  \caption*{\footnotesize{\textit{Note}: Posterior distribution of impulse responses in percent. The median is depicted in grey, the 16$^{\text{th}}$ and 84$^{\text{th}}$ percentile are indicated by the light grey area. The dashed line indicates zero. Country codes may be found in \cref{app:data-app}.}}
	\label{irf_factor_exintra}
\end{figure}

\begin{figure}[!htbp]
	\includegraphics[width=1\textwidth]{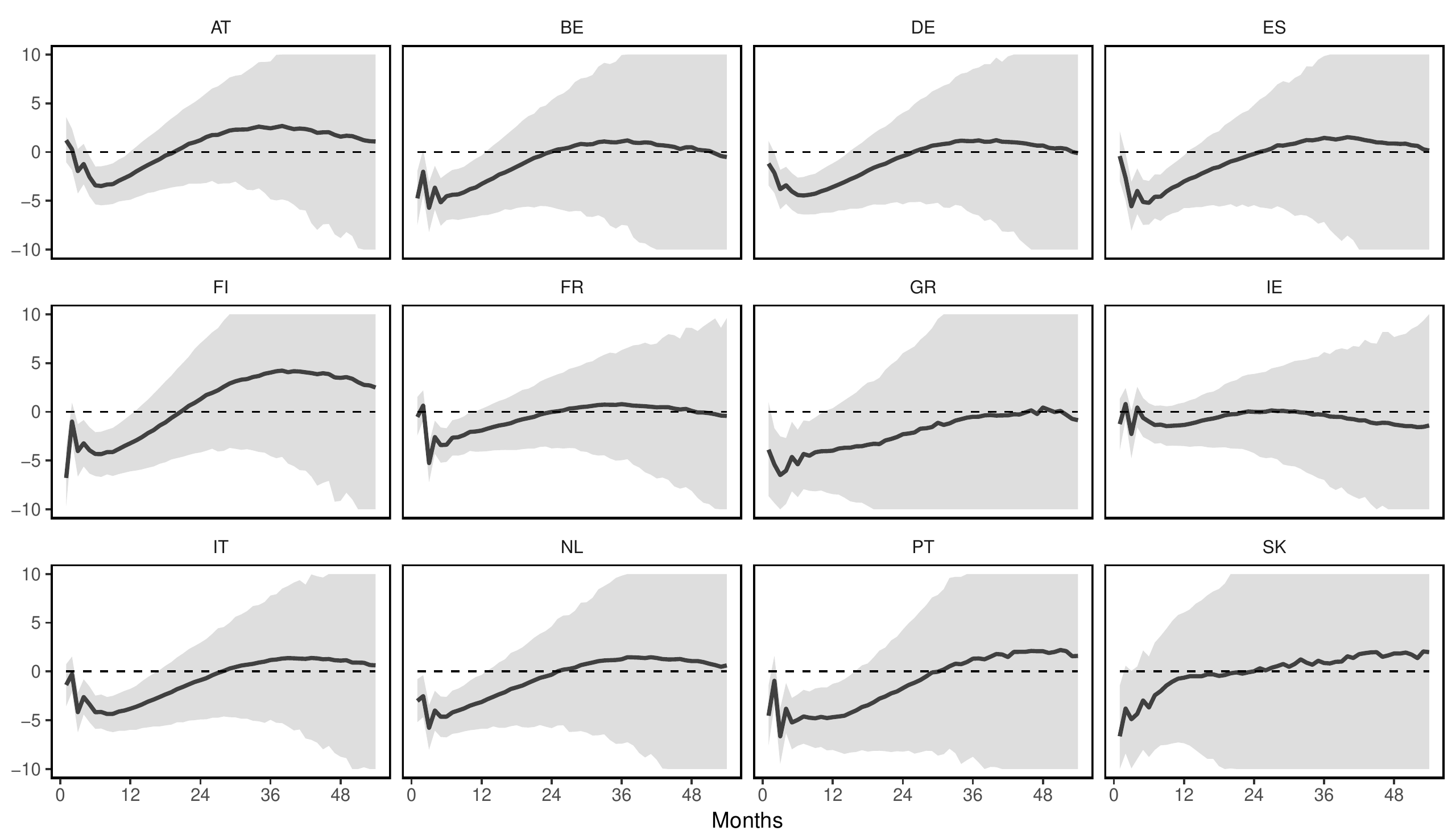}
	\caption{Impulse response functions of exports outbound from EU countries.}
  \caption*{\footnotesize{\textit{Note}: Posterior distribution of impulse responses in percent. The median is depicted in grey, the 16$^{\text{th}}$ and 84$^{\text{th}}$ percentile are indicated by the light grey area. The dashed line indicates zero. Country codes may be found in \cref{app:data-app}.}}
	\label{irf_factor_exextra}
\end{figure}

\Cref{irf_factor_exintra} exhibits the impact of an uncertainty shock on exports within the EU.\footnote{Note that most EU countries trade more than 80 percent of their entire goods and services within the EU.} On impact, the effect is modest, but reaches a maximum of an approximate five percent decline after six to eight months. This suggests a tight link between (intra-EU) trade and European uncertainty. The drop in within-EU exports is remarkably long lasting. After 12 to 15 months, however, exports begin to increase. We do not observe a significant rebound effect, even though substantial mass of the posterior distribution is shifted away from zero. The movements are almost identical for all countries considered, but vary slightly in magnitude and duration. The strongest effects are observed in Finland, Italy and the Netherlands. Comparatively smaller effects occur in Ireland. This may be due to the special role of Ireland within the EU being home to headquarters of multinationals. Considering impulse responses with respect to exports outbound from the EU, we observe a similar pattern. \Cref{irf_factor_exextra} shows a significant drop after a few months, dying out in most countries after roughly a year.

To assess the robustness of our results, we estimate various other specifications. One alternative is based on a subset of countries with the highest GDP in the Euro area (Germany, Spain, France, Italy and the Netherlands). Hereby, we observe that the latent common factor in the error term is robust to differing selections. Considering the impulse responses in this specification, we find slightly different effects compared to the bigger model. However, the sign of the responses and thus the derived implications hold. Different ordering of the variables did not result in significantly different findings.

\section{Concluding Remarks}
\label{sec:conclusion}
Articles on estimating impacts of uncertainty and macroeconomic consequences jointly in a multi-country setting are relatively sparse. We contribute to this literature by analyzing the special case of a shock to the Euro area. In this paper, we estimate a Bayesian vector autoregressive model with a factor stochastic volatility structure in the error term. The econometric modeling approach employed in this article closely resembles the work by \citet{crespo2017macroeconomic}. The results are in line with recent findings. Treating uncertainty as a latent quantity allows comparisons with proxies such as the EuroVIX or business and consumer confidence indices. We find that the estimated quantity closely tracks these measures, covering financial volatility and macroeconomic uncertainty.

Before considering the impacts of second-moment shocks on macroeconomic quantities, we indicate shares of innovation variance explained by the uncertainty factor and find that it contributes substantially. Regarding impulse response analysis of the variables in the system to an uncertainty shock, we obtain significant results of a decrease in real activity in most Euro area countries over a period of roughly one year. Furthermore, we find significant effects of uncertainty considering rises in unemployment, decreases in the short-term interest rate, equity prices, as well as intra Euro area exports and exports to non-EU countries.

Greece, Ireland and Slovakia are often affected slightly differently by the shock than other countries. Three distinct objectives may be identified for future considerations. First, additional factors of uncertainty may be used to differentiate between different kinds of uncertainty, such as financial volatility, fiscal or monetary policy uncertainty. Second, depending on interdependencies of economies and the magnitude of business cycle synchronization, clusters of countries and cluster-specific reactions might be of interest. Finally, reflecting the notion of higher impacts of uncertainty during recessions, non-linear time series models might be an interesting avenue for future research.

\small\singlespacing
\bibliographystyle{./bibtex/cit_econometrica}
\bibliography{./bibtex/lit}
\addcontentsline{toc}{section}{References}

\onehalfspacing\normalsize

\newpage
\begin{appendices}\crefalias{section}{appsec}
\setcounter{equation}{0}
\renewcommand\theequation{A.\arabic{equation}}

\section{Prior hyperparameter values}
\label{app:hyperparameters}
First, we choose $\kappa_1 = 0.6$. For higher order lags, we use $\kappa_p=0.6/p^2$. Second, we choose $c_j=3$ and $d_j = 0.03$. Third, regarding the elements of the factor loadings matrix $\bm{X}$, we assume a relatively uninformative prior structure, where each element comes frome a zero mean Gaussian distribution with a prior variance of ten.

 For the unconditional mean, the persistence parameter of the log-volatilities and the innovation variance of the log-volatilities we have for $s \in \{h,\omega\}$
\begin{equation}
\begin{aligned}
    \mu_i^{(s)} &\sim \N(0,10)\\
    \frac{\phi_i^{(s)}+1}{2} &\sim B(5,1.5)\\
    \Xi^{(s)}_i &\sim G\left(\frac{1}{2},\frac{1}{2}\right).
\end{aligned}
\end{equation}

\section{Data}
\label{app:data-app}
The variables are grouped as follows: Euro Area Aggregate (EA19), Austria (AT), Belgium (BE), Germany (DE), Greece (GR), Spain (ES), Finland (FI), France (FR), Ireland (IE), Italy (IT), the Netherlands (NL), Portugal (PT), Slovakia (SK) and exogenous variables (exo).

\begin{table*}[!htbp]
\caption{List of variables and corresponding source.}
\small
\begin{center}
\begin{threeparttable}
\begin{tabular*}{\textwidth}{@{\extracolsep{\fill}} l l l l c}
\toprule
\textbf{Variable} & \multicolumn{1}{l}{\textbf{Description}} & \multicolumn{1}{c}{\textbf{Transformation}} & \multicolumn{1}{c}{\textbf{Source}} & \multicolumn{1}{c}{\textbf{Comments}}\\
\midrule
y & Real GDP & Logs & OeNB & --- \\ 
p & Consumer price index & LogDiffs & OeNB & --- \\ 
stir & Short-term interest rate & --- & OeNB & --- \\ 
eq & EUROSTOXX Main index & Logs & stoxx.com & --- \\ 
cci & Consumer confidence index & Logs & OECD & EA19 specific \\ 
bci & Business confidence index & Logs & OECD & EA19 specific \\ 
unemp.r & Unemployment rate & Season. adjusted & Eurostat & --- \\ 
ciss & Stress index$^{*}$ & Logs & ECB & EA19 specific \\ 
ex.intra & Intra EU exports & Deseason. logs & Eurostat/Comext & in mio. EUR \\ 
ex.extra & non-EU exports & Deseason. logs & Eurostat/Comext & in mio. EUR \\ 
poil & Oil price index & Logs & OeNB & exogenous \\ 
eurovix & EUROSTOXX volatility index & Logs & stoxx.com & EA19 specific \\[1.2ex]
\bottomrule
\end{tabular*}
\begin{tablenotes}[para,flushleft]
\footnotesize{\textit{Note}: ECB -- European Central Bank; OeNB -- Oesterreichische Nationalbank; OECD -- Organisation for Economic Co-operation and Development; $^{*}$Composite Indicator of Systemic Stress. Logs -- Natural Logarithm, Diffs -- Differences.}
\end{tablenotes}
\end{threeparttable}
\end{center}
\label{a}
\end{table*}

\end{appendices}

\end{document}